\documentclass[runningheads]{llncs}
\usepackage{graphicx}
\usepackage{subfigure}
\usepackage{amsmath}
\usepackage{amssymb}
\usepackage{wrapfig}
\usepackage{upgreek}
\usepackage{booktabs}

\newcommand{\argmax}[1]{\underset{#1}{\operatorname{arg}\,\operatorname{max}}\;}

\begin{document}
\title{Longitudinal Correlation Analysis \\for Decoding Multi-Modal Brain Development}
\titlerunning{Longitudinal Correlation Analysis}
%
\author{Qingyu Zhao\inst{1}, 
Ehsan Adeli\inst{1,2}, 
Kilian M. Pohl\inst{1,3}}
\institute{School of Medicine, Stanford University, Stanford, USA \and Computer Science Department, Stanford University, Stanford, USA \and Center of Health Sciences, SRI International, Menlo Park, USA}

\maketitle              
\begin{abstract} 
Starting from childhood, the human brain restructures and rewires throughout life. Characterizing such complex brain development requires effective analysis of longitudinal and multi-modal neuroimaging data. Here, we propose such an analysis approach named Longitudinal Correlation Analysis (LCA). LCA couples the data of two modalities by first reducing the input from each modality to a latent representation based on autoencoders. A self-supervised strategy then relates the two latent spaces by jointly disentangling two directions, one in each space, such that the longitudinal changes in latent representations along those directions are maximally correlated between modalities. We applied LCA to analyze the longitudinal T1-weighted and diffusion-weighted MRIs of 679 youths from the National Consortium on Alcohol and Neurodevelopment in Adolescence. Unlike existing approaches that focus on either cross-sectional or single-modal modeling, LCA successfully unraveled coupled macrostructural and microstructural brain development from morphological and diffusivity features extracted from the data. A retesting of LCA on raw 3D image volumes of those subjects successfully replicated the findings from the feature-based analysis. Lastly, the developmental effects revealed by LCA were inline with the current understanding of maturational patterns of the adolescent brain.

\end{abstract}
\section{Introduction}
The human brain undergoes profound changes over the entire life span \cite{Petrican2017}. Starting from childhood, gray matter volume declines \cite{Giedd2004}, in part reflecting synaptic pruning \cite{Natu2019}, while white matter volume increases, contributing to efficient neural signaling and transmission \cite{Giedd2004}. The marked morphological development, along with the maturation in microstructural fiber pathways \cite{Lebel2008}, gives rise to the remodeling of functional brain circuits, supporting the capacity for high-level cognition \cite{Simmonds2013}. To advance the understanding of such multifaceted neurodevelopment, studies have increasingly relied on analyzing \textit{longitudinal} and \textit{multi-modal} MRI data collected over specific age ranges \cite{Sadeghi2010}.

While deep learning techniques have led to tremendous progress in the analysis of brain MRIs \cite{Zhu2019}, the focus has been on either cross-sectional or single modality data. For example, existing longitudinal models for assessing neurodevelopment are primarily  designed for structural MRIs only \cite{zhao2020lssl}.
Multi-modal analysis \cite{Chen2019,Han20,he2020momentum,tsai2021multi}, on the other hand, explores common factors across modalities using cross-sectional data, so the identified factors are not guaranteed to characterize development. Another limitation of deep learning in analyzing development is that they are often formulated as supervised learning models with respect to pseudo-marker of developmental stages, such as age. As suggested in \cite{zhao2020lssl}, such age-based supervision is sub-optimal as brain development is highly heterogeneous in populations with similar chronological ages.

To identify developmental effects from two longitudinal MRI modalities without the guidance of age, we propose a longitudinally self-supervised deep learning approach, named Longitudinal Correlation Analysis (LCA). LCA consists of an autoencoding structure that separately reduces the data of either modality to a latent representation. Motivated by Canonical Correlation Analysis (CCA) \cite{Hardle2007}, LCA then relates the two latent spaces by jointly estimating one direction in each space, such that the longitudinal changes (of latent representations) along those directions are maximally correlated between modalities. Training of LCA is self-supervised as it only requires grouping the repeated measures of each individual but does not rely on chronological age (or any other supervisory signal). 

We applied LCA to analyze the longitudinal multi-modal data of 679 subjects from the National Consortium on Alcohol and Neurodevelopment in Adolescence (NCANDA) \cite{brown2015national} to investigate the macrostructural and microstructural brain development of adolescents. Based on features extracted from longitudinal T1 and diffusion-weighted MRIs, LCA successfully revealed coupled developmental effects from the two modalities while baseline methods failed to do so. We then successfully replicated the findings of this feature-based analysis by retesting LCA on the raw longitudinal T1 images and Fractional Anisotropy maps. Lastly, we show that the developmental effects revealed by LCA were inline with the current understanding of maturational patterns of the adolescent brain. 

\section{Method}
We first review the Canonical Correlation Analysis (CCA) \cite{Hardle2007}  for identifying association across two modalities. We then discuss the difficulties of linking the CCA solution to interpretable factors. This limitation motivates our design of self-supervised learning based on the repeated measures of longitudinal data to jointly disentangle developmental effects across modalities.

\noindent\textbf{Canonical Correlation Analysis.}
Let $X:= [\mathbf{x}^1, \cdots, \mathbf{x}^S] \in \mathbb{R}^{n \times S}$ be the data of the first modality from $S$ subjects, where $\mathbf{x}^s \in \mathbb{R}^n$ is the data of subject `$s$'. Likewise, let $Y := [\mathbf{y}^1, \cdots, \mathbf{y}^S] \in \mathbb{R}^{m \times S}$ be the data of the second modality. CCA aims to find two vectors $\boldsymbol{\tau}_x \in \mathbb{R}^n$ and $\boldsymbol{\tau}_y \in \mathbb{R}^m$, such that the projections of $X$ and $Y$ onto these vectors are maximally correlated, i.e., 
\begin{equation}
     (\boldsymbol{\tau}_x,\boldsymbol{\tau}_y) := \argmax{\boldsymbol{\tau}'_x,\boldsymbol{\tau}'_y} \: \text{corr}({\boldsymbol{\tau}'_x}^\top X,{\boldsymbol{\tau}'_y}^\top Y).
    \label{eq:cca}
\end{equation}
We call the optimal $\boldsymbol{\tau}_x$ and $\boldsymbol{\tau}_y$ the canonical vectors and the resulting projections $p_x := \boldsymbol{\tau}_x^\top X$ and $p_y := \boldsymbol{\tau}_y^\top Y$ the {\it canonical variables}. 

\begin{figure}[!t]
    \centering
    \includegraphics[trim=80 230 380 140, clip,width=0.90\linewidth]{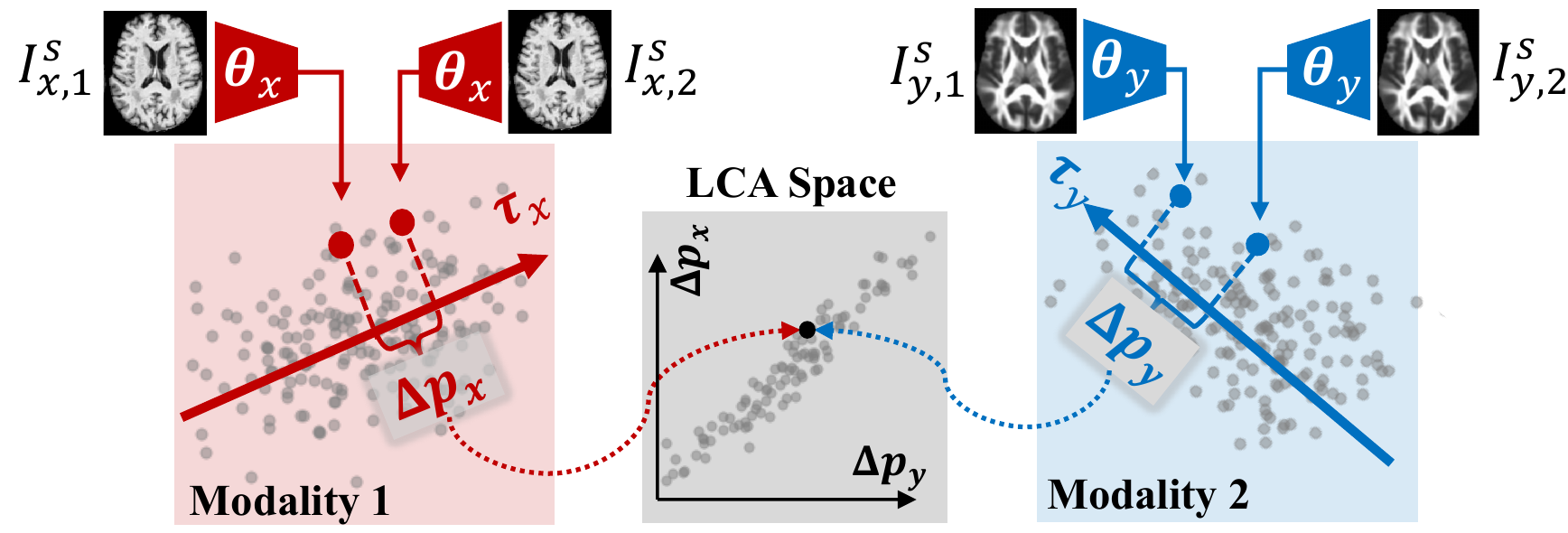}
    \caption{LCA finds two directions $\boldsymbol{\tau}_x$ and $\boldsymbol{\tau}_y$ in two latent spaces, one for each modality, such that the projections ($\Delta p_x$, $\Delta p_y$) of longitudinal changes are maximally correlated.}
    \label{fig:my_label}
\end{figure}

While these canonical variables characterize the strongest association across modalities, their meanings are not easily interpretable as they cannot be exactly linked to any single real-life factor underlying the data (such as age or gender). To conceptually illustrate this ambiguity problem, 
we assume that the data are generated by $K$ hidden factors $\boldsymbol{\alpha} := [\alpha^1,\cdots,\alpha^K]^\top$ through linear functions. In other words, $\mathbf{x} = U\boldsymbol{\alpha}$ and $\mathbf{y} = V\boldsymbol{\alpha}$ with $U := [\mathbf{u}^1,\cdots,\mathbf{u}^K]\in \mathbb{R}^{n\times K}$ and $V := [\mathbf{v}^1,\cdots,\mathbf{v}^K]\in \mathbb{R}^{m\times K}$ consisting of $K$ components. 
Now, we define the canonical variables $p_x$ and $p_y$ as being linked to a single factor $\alpha_k$ if $\boldsymbol{\tau}_x$ and $\boldsymbol{\tau}_y$ are orthogonal to all components except for $\mathbf{u}^k$ and $\mathbf{v}^k$, such that
\begin{equation}
 p_x = \boldsymbol{\tau}_x^\top \mathbf{x} = \boldsymbol{\tau}_x^\top U \boldsymbol{\alpha} = \boldsymbol{\tau}_x^\top \mathbf{u}^k \alpha^k \propto \alpha^k \text{, and likewise,} \; p_y  \propto \alpha^k.
    \label{eq:disentanglement}
\end{equation}
In practice, the above condition cannot be guaranteed in the CCA setting as $\boldsymbol{\alpha}$ cannot be uniquely determined. In fact, for any orthogonal transformation $R$, one can re-define the generative procedure with respect to $\boldsymbol{\alpha}^\prime=R\boldsymbol{\alpha}$ as
\begin{equation}
    \mathbf{x} = U\boldsymbol{\alpha} = UR^\top R\boldsymbol{\alpha} = U^\prime\boldsymbol{\alpha}^\prime \text{, and likewise,} \; \mathbf{y} = V^\prime\boldsymbol{\alpha}^\prime.
\end{equation}
This indicates the same $p_x$ and $p_y$ are linked to a family of possible factors defined by the quotient group of orthogonal transformation. 
In fact, such ambiguity is not limited to CCA but generally exists in probabilistic formulations of independent component analysis (ICA) and principal component analysis (PCA),  where the generative factors can only be determined up to rotation \cite{locatello2018challenging}. 

\noindent\textbf{Longitudinal Correlation Analysis (LCA).} 
We now show that we can leverage longitudinal data to relate canonical variables to the single factor of brain development without ambiguity by assessing the `differential effects' between repeated measures. 
To do this, we first observe that while $\mathbf{x}$ is a composite of multiple factors, the partial derivative $\partial  \mathbf{x} /\partial \alpha^k$ is a single component $\mathbf{u}^k$. In other words, if we perturb $\alpha^k$ by $\Delta \alpha^k$ while fixing the value of other factors, the resulting difference in $\mathbf{x}$ is simply $\Delta \mathbf{x} = \mathbf{u}^k\Delta \alpha^k$, which is solely dependent on the $k^{th}$ component. Moreover, the resulting difference in canonical variables are also guaranteed to link to one component because
\begin{equation}
        \Delta p_x = \boldsymbol{\tau}_x^\top \Delta \mathbf{x}  = (\boldsymbol{\tau}_x^\top \mathbf{u}^k) \Delta \alpha^k \propto \Delta \alpha^k \text{, and likewise,} \;  \Delta p_y \propto \Delta \alpha^k.
    \label{eq:pd}
\end{equation}

This observation implies that we can transfer the CCA setup on $\mathbf{x}$ and $\mathbf{y}$ to a setup on the differential vectors $\Delta \mathbf{x}$ and $\Delta \mathbf{y}$ to focus on finding cross-modal information related to one single factor. A definition of $\Delta \mathbf{x}$ and $\Delta \mathbf{y}$ that well fits with longitudinal studies is the difference between the repeated measures of an individual. Unlike the unknown factors underlying $\mathbf{x}$ and $\mathbf{y}$, these differential vectors solely encode a known factor of brain development as all other factors such as race and gender are static across visits (assuming no brain damaging events occur between visits). LCA then maximizes the correlation between $\Delta p_x$ and $\Delta p_y$ to unravel coupled developmental effects across modalities.

We embed LCA in an autoencoder setting. Let $I_{x,1}^s$ and $I_{x,2}^s$ be two images (or feature vectors) of the first modality from subject $s$ with $I_{x,2}^s$ scanned after $I_{x,1}^s$. We assume the images can be reduced to low-dimensional latent representations by an encoder network with parameters $\theta_x$, i.e., $\mathbf{x}^s_1:=f(I_{x,1}^s;\theta_x)$ and $\mathbf{x}^s_2:=f(I_{x,2}^s;\theta_x)$. The decoder network $g$ with parameters $\phi_x$ can then reconstruct the latent representations. Let $\Delta \mathbf{x}^s = \mathbf{x}^s_2 - \mathbf{x}^s_1$, $\Delta 
\mathbf{y}^s = \mathbf{y}^s_2 - \mathbf{y}^s_1$ and let $\Delta X(\theta_x) :=[\Delta \mathbf{x}^1,\cdots,\Delta \mathbf{x}^s]$, $\Delta Y(\theta_y):= [\Delta \mathbf{y}^1,\cdots,\Delta \mathbf{y}^s]$. We now propose to couple the two developmental effects by finding $\boldsymbol{\tau}_x$ and $\boldsymbol{\tau}_y$ such that projections of the longitudinal changes are maximally correlated across the group while reducing the reconstruction loss based on the mean-squared error $||\cdot||^2$ :
\begin{align}
\label{eq:lca}
\argmax{\boldsymbol{\tau}_x,\boldsymbol{\tau}_y,\theta_x,\theta_y,\phi_x,\phi_y}& \: \text{corr}(\boldsymbol{\tau}_x^\top \Delta X(\theta_x),\boldsymbol{\tau}_y^\top \Delta Y(\theta_y)) - \\
&\lambda\sum_{s=1}^S\sum_{t=1}^2\left(||g(f(I_{x,t}^s;\theta_x);\phi_x),I_{x,t}^s||^2+||g(f(I_{y,t}^s;\theta_y);\phi_y),I_{y,t}^s||^2\right) \nonumber 
\end{align}
This optimization is fully self-supervised as we only need to pair the repeated measures within each individual. Although the training only correlates $\Delta p_x$ and $\Delta p_y$, the actual values of $p_x$ and $p_y$ can be directly computed by projecting data onto $\boldsymbol{\tau}_x$ and $\boldsymbol{\tau}_x$ during validation.

\section{Experiments}
\subsection{Setup}
We investigate adolescent brain development by applying LCA to feature vectors or 3D volumes derived from T1-weighted (T1) MRIs and from 3D diffusivity maps associated with diffusion weighted MRIs (DWI).

\textbf{Data.} The testing data set consisted of the T1 MRI and DWI of 2169 visits that were from 679 healthy youths recruited by NCANDA (no more than 6 annual visits and in  average 3.2 visits per subject; age at baseline visits: 15.7$\pm$2.4; 340 boys/339 girls) who met the no-to-low drinking criteria based on the adjusted Cahalan score \cite{pfefferbaum2018altered}. For each subject, we constructed pairs from two different visits and randomly selected 3 visit pairs for subjects with more than 3 visits to avoid over-emphasizing those subjects.  This procedure resulted in 1054 visit pairs with the average age difference between paired visits being 2.7$\pm$0.9 years.

\textbf{Preprocessing.} T1 MRI were preprocessed by denoising, bias-field correction, and skull striping. Cortical thickness and surface area of 68 bilateral cortical regions and volume of 31 sub-cortical regions were extracted by FreeSurfer 5.3 \cite{Fischl2012} resulting in a 99 dimensional feature for each image. Then the skull-stripped images were further affinely registered to the SRI24 atlas \cite{Rohlfing2008} and reduced to $64\times64\times64$ resolution. DWIs were first registered to the structural images and skull stripped. Preprocessing continued with bad single shots removal, echo-planar, and Eddy-current distortion correction. Camino \cite{Cook2005} computed diffusivity maps including fractional anisotropy (FA), medium diffusivity (MD), L1, and LT maps. Average regional measures were extracted from each map by reducing the map to a skeleton via Tract-Based Spatial Statistics (TBSS) \cite{Smith2006tbss}, dividing the skeleton into  27 regions according to the Johns Hopkins University (JHU) DTI atlas \cite{Mori2005}, and averaging the skeleton values for each region. This resulted in a 108 dimensional feature. In parallel, the 3D FA maps were reduced to $64\times64\times64$ resolution. 

\textbf{Implementation.} For feature-based analysis, the encoder was a two-layer perceptron with dimension (64, 32) with tanh activation. The decoder adopted the inverse structure. For 3D-image-based analysis, the encoder was an encoder composed of 4 stacks of $3\times3\times3$ convolution/ReLU/max-pooling layers with feature channel (16, 32, 64, 16). Then, a dense layer reduced the output to a 512 dimensional representation. We set a non-informative $\lambda=1/(2S)$. Models were implemented in Keras 2.2.2 (see code at \url{https://github.com/QingyuZhao/LCA}) and run on an Nvidia Quadro P6000 GPU. Training was confined to the 1054 visit pairs and performed by the Adam optimizer with a 0.0002 learning rate for 200 epochs.

\textbf{Evaluation.} We applied the trained model to derive the canonical variables $p_{\text{T1}}$ and $p_{\text{DWI}}$ of all 2169 visits (i.e., including also visits that were not paired during training). We then used two types of statistical analysis to examine the relationship between the canonical variables with age (we regard age only as an approximate marker for quantifying brain development but not the ground-truth for $p_{\text{T1}}$ and $p_{\text{DWI}}$). To remove the impact of repeated measures, we first computed the Pearson's correlation \textbf{$r$} between $p_{\text{T1}}$ at the baseline visit of each subject and their age at baseline. To account for the consistency of $p_{\text{T1}}$ across the repeated measures, we then parameterized a linear mixed effect (LME) model, which estimated a cubic group-level trajectory of $p_{\text{T1}}$ over age with each subject having a random intercept. The goodness of fit of LME was examined by the Akaike information criterion (AIC) \cite{Aho2014}. Standard deviation of these statistics was generated based on a bootstrapping procedure, which repeated the training and validation 10 times based on resampling for each subject the visit pairs used for  training. Lastly, the age-related statistics were also generated for $p_{\text{DWI}}$. 

\textbf{Baselines.} While many existing models are designed to leverage association between multi-view data \cite{Han20,he2020momentum,tsai2021multi}, to the best of our knowledge, there are no unsupervised or self-supervised multi-view methods that also disentangle latent directions from longitudinal data. We, therefore, compared LCA to multi-modal cross-sectional methods, including Deep CCA (DCCA) \cite{Andrew2013dcca} and Deep Canonical Correlation Autoencoders (DCCAE) \cite{wang2016deep}, which were extensions of CCA in deep learning settings. All 2169 T1 and DWI images were used for training these two methods as they did not rely on repeated measures. We then further compared LCA to Longitudinal Self-Supervised Learning (LSSL) \cite{zhao2020lssl}, a longitudinal method that estimates a brain aging direction in the latent space of a single modality. Training LSSL was based on the same set of visit pairs but was independently performed on either modality.

\subsection{Results}
\begin{figure}[!t]
    \centering
    \includegraphics[trim=114 250 170 100, clip,width=\linewidth]{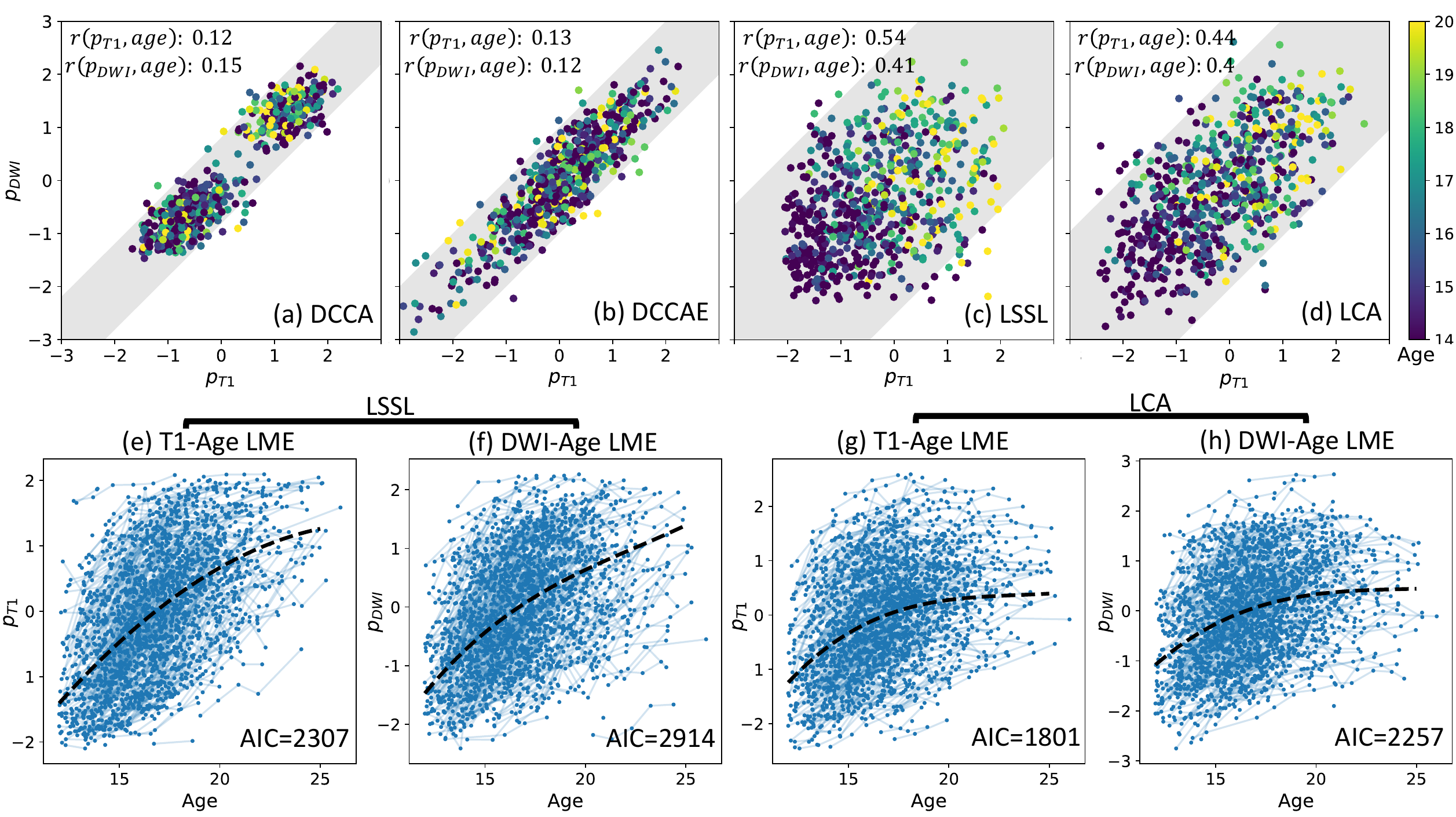}
    \caption{Feature-based analysis. (a)-(d): canonical variables at the baseline visits of subjects for the 4 comparison methods. Each point is color-coded by age. Width of the gray bands corresponds to Pearson's correlation  $r(p_{\text{T1}},p_{\text{DWI}})$; (e)-(f): LME fitting between age and $p_{\text{T1}}$, $p_{\text{DWI}}$ for LSSL and LCA. Black curves are the average estimation of the cubic fixed effects by bootstrapping.}
    \label{fig:feature}
\end{figure}

\begin{figure}[!t]
    \centering
    \includegraphics[trim=130 260 55 240, clip,width=\linewidth]{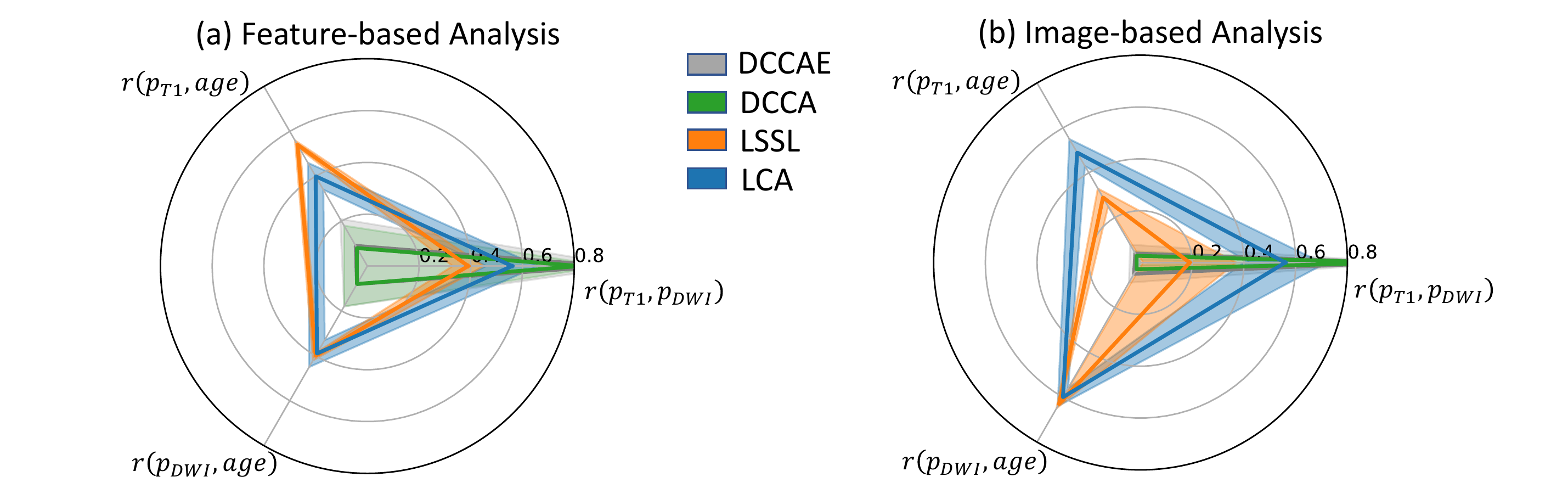}
    \caption{Mutual correlation between age and the canonical variables $p_{\text{T1}}$, $p_{\text{DWI}}$ for feature-based and image-based experiments. Width of triangle edges correspond to standard deviation of the correlation derived by bootstrapping. }
    \label{fig:radar}
\end{figure}

\textbf{Feature-based Analysis.} Figs. \ref{fig:feature}(a)-(d) display the canonical variables associated with the two modalities of the 679 subjects at their baseline visits. The gray band in each figure corresponds to the Pearson's correlation $r(p_{\text{T1}},p_{\text{DWI}})$. 
Unlike the baselines, LCA resulted in highly correlated $p_{\text{T1}}$ and $p_{\text{DWI}}$, both of which also significantly correlated with age (see also Fig. \ref{fig:radar}(a)). Note, the significant group-level correlation between $p_{\text{T1}}$ and $p_{\text{DWI}}$ ($r=0.61$, Fig. \ref{fig:feature}d) were the result of training LCA  solely on longitudinal differences $\Delta p_{\text{T1}}$ and $\Delta p_{\text{DWI}}$. Moreover, both canonical variables highly correlated with age ($r\geq0.4$), indicated by the smooth color transition along the diagonal direction (older youths were color-coded as yellow). Again, this group-level correlation within the whole age span from 12 to 25 years was learned from pairs of visits, whose average time between visits was 2.7 years. This indicates that LCA successfully disentangled the factor linked to brain development within the multi-modal data. In comparison, the first pair of canonical variables identified by DCCA and DCCAE had higher correlation ($r(p_{\text{T1}},p_{\text{DWI}})=0.87$), but neither variable was related to brain development, indicated by the substantially lower correlation with age ($r\leq0.15$, Fig. \ref{fig:radar}a) and the less pronounced color transition in Figs. \ref{fig:feature}(a)+(b). In fact, age correlation for all 32 pairs of canonical variables of DCCA and DCCAE were relatively low ($r=0.08\pm0.06$), which supports our speculation that each canonical variable of conventional CCA is a composite of multiple factors. While the single-modal analysis of LSSL could separately reveal developmental effect in each modality, the identified $p_{\text{T1}}$ and $p_{\text{DWI}}$ showed lower correlation ($r=0.39$, Fig. \ref{fig:feature}c) than LCA. The differences between modality-specific LSSL results are also evident in Fig. \ref{fig:radar}a, where the developmental effect was more pronounced in T1 features ($r(p_{\text{T1}},age)=0.54$) than in DWI features ($r(p_{\text{DWI}},age)=0.40$). This was not the case for LCA, which resulted in balanced age correlation across modalities (Fig. \ref{fig:radar}a).

\begin{figure}[!t]
    \centering
    \includegraphics[trim=30 190 60 100, clip,width=0.95\linewidth]{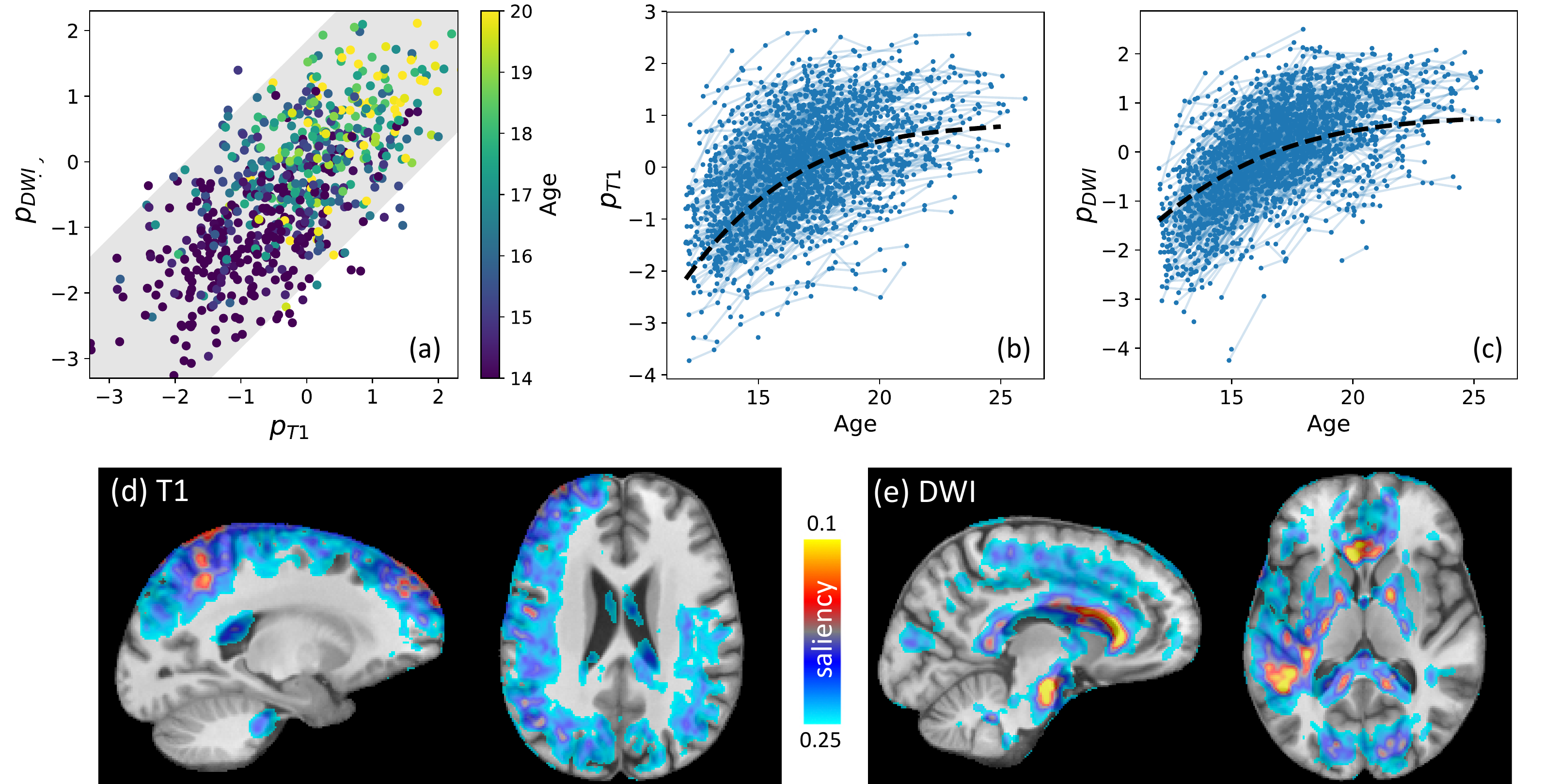}
    \caption{Image-based analysis by LCA: (a) $p_{\text{T1}}$, $p_{\text{DWI}}$ at the baseline visits of subjects; (b)-(c): LME fitting on age; (d)-(e): Saliency maps overlaid on the structural atlas indicating regions extracted from T1 and DWI that show coupled development. }
    \label{fig:img}
\end{figure}
Fig. \ref{fig:feature}(e)-(h) shows individual developmental trajectories (blue) estimated by LSSL and LCA and the group-level trajectories derived by the post-hoc LME fitting. Despite both methods producing strong age correlation, the developmental trajectories of $p_{\text{T1}}$ and $p_{\text{DWI}}$ revealed by LCA were more robust and realistic than those by LSSL. According to the LME fitting (Figs. \ref{fig:feature}(g)+(h)), LSSL resulted in lower AIC metrics than LCA, suggesting the individual trajectories estimated by LCA were more consistent with the group-level trajectory. Moreover, LCA revealed that both macrostructural and microstructural development was fast in the teen years but plateaued during adulthood (after age 20), which comported with maturational patterns reported for the NCANDA cohort  \cite{pohl2016,pfefferbaum2018altered}. In comparison, $p_{\text{T1}}$ and $p_{\text{DWI}}$ estimated by LSSL showed significant increase even after reaching adulthood (Figs. \ref{fig:feature}(e)+(f)), which might be due to overfitting. With respect to LCA, the variance in $p_{\text{T1}}$ or $p_{\text{DWI}}$ at any given age potentially indicates that the brain developmental stage is highly heterogeneous during adolescence, which supports our claim that self-supervised learning is potentially a better approach for quantifying that stage compared to age-guided supervision. 

\textbf{Image-based Analysis.} Similar to the previous experiment, DCCA and DCCAE revealed strongly coupled factors that did not reflect brain development (Fig. \ref{fig:radar}b). Unlike the previous experiment, LSSL revealed pronounced developmental effects only in DWI but resulted in significantly lower correlation (than the feature-based analysis) between age and $p_{\text{T1}}$. As a result, we observed dissociation between $p_{\text{T1}}$ and $p_{\text{DWI}}$ by LSSL ($r(p_{\text{T1}},p_{\text{DWI}})=0.13$, Fig. \ref{fig:radar}b) compared to the feature-based result. 
On the other hand, the results of LCA were inline with the findings from the previous experiment. The mutual correlation among $p_{\text{T1}}$, $p_{\text{DWI}}$, and age all increased compared to the feature-based results, suggesting that the raw imaging data contained more comprehensive information compared to low-dimensional features. The three correlation values were well balanced (Fig. \ref{fig:radar}b), suggesting that joint self-supervised learning might regularize the disentanglement across modalities.

Lastly, we visualized regions that were associated with development in either modality. We did so by quantifying the voxelwise importance in driving the change in $p_{\text{T1}}$ and $p_{\text{DWI}}$ via guided saliency map \cite{simonyan2014deep} being applied to each T1 MRI and FA map. The average saliency maps across the cohort (Figs.  \ref{fig:img}(d)-(e)) highlighted gray matter areas from the T1 images and highlighted white matter pathways (including the corpus callosum and brainstem) in the FA maps. Such distinction in regions was to be expected as gray matter morphology was only visible in structural images, whereas FA maps were specific to the quantification of white matter integrity. This supports that LCA successfully revealed complementary aspects for adolescent brain maturation from multi-modal data.

\section{Conclusion}
We have proposed a self-supervised multi-modal data analysis approach named Longitudinal Correlation Analysis. By coupling the developmental effects of repeated measures in latent spaces, LCA successfully uncovered the multifaceted developmental effects during adolescence from T1 and diffusion-weighted MRI. Reproducibility between feature-based and image-based analyses highlighted the robustness of the method. The correlation-based self-supervision may have a broader impact for multi-view and contrastive learning, applications of which remained to be explored in the future.

\textbf{Acknowledgment.} This  research  was  supported  in  part  by  NIH  
U24 AA021697 and Stanford HAI AWS Cloud Credit. The data were part of the public NCANDA data release NCANDA\_PUBLIC\_5Y\_REDCAP\_V01, NCANDA\_PUBLIC\_5Y\_STRUCTURAL\_V01,  NCANDA\_PUBLIC\_5Y\_DIFFUSION\_V01 \cite{ncanda_data_release}, whose collection and distribution were supported by NIH funding AA021697, AA021695, AA021692, AA021696, AA021681, AA021690, and AA02169.


%
%
%
\bibliographystyle{splncs}
\bibliography{miccai}

\end{document}